\def \SAIT #1 #2 {{\em Mem.\ Soc.\ Astron.\ It.\/} {\bf #1}, #2}
\def \MESS #1 #2 {{\em The Messenger\/} {\bf #1}, #2}
\def \ASTRNACH #1 #2 {{\em Astron. Nach.\/} {\bf #1}, #2}
\def \AAP #1 #2 {{\em Astron. Astrophys.\/} {\bf #1}, #2}
\def \AAL #1 #2 {{\em Astron. Astrophys. Lett.\/} {\bf #1}, L#2}
\def \AAR #1 #2 {{\em Astron. Astrophys. Rev.\/} {\bf #1}, #2}
\def \AAS #1 #2 {{\em Astron. Astrophys. Suppl. Ser.\/} {\bf #1}, #2}
\def \AJ #1 #2 {{\em Astron. J.\/} {\bf #1}, #2}
\def \ANNREV #1 #2 {{\em Ann. Rev. Astron. Astrophys.\/} {\bf #1}, #2}
\def \APJ #1 #2 {{\em Astrophys. J.\/} {\bf #1}, #2}
\def \APJL #1 #2 {{\em Astrophys. J. Lett.\/} {\bf #1}, L#2}
\def \APJS #1 #2 {{\em Astrophys. J. Suppl.\/} {\bf #1}, #2}
\def \APSS #1 #2 {{\em Astrophys. Space Sci.\/} {\bf #1}, #2}
\def \ASR #1 #2 {{\em Adv. Space Res.\/} {\bf #1}, #2}
\def \BAIC #1 #2 {{\em Bull. Astron. Inst. Czechosl.\/} {\bf #1}, #2}
\def \JSQRT #1 #2 {{\em J. Quant. Spectrosc. Radiat. Transfer\/} {\bf #1}, #2}
\def \MN #1 #2 {{\em Mon. Not. R. Astr. Soc.\/} {\bf #1}, #2}
\def \MEM #1 #2 {{\em Mem. R. Astr. Soc.\/} {\bf #1}, #2}
\def \PLR #1 #2 {{\em Phys. Lett. Rev.\/} {\bf #1}, #2}
\def \PASJ #1 #2 {{\em Publ. Astron. Soc. Japan\/} {\bf #1}, #2}
\def \PASP #1 #2 {{\em Publ. Astr. Soc. Pacific\/} {\bf #1}, #2}
\def \NAT #1 #2 {{\em Nature\/} {\bf #1}, #2}

\documentstyle{memsait}
\input epsf.sty
\begin{opening}
\title{MAPS OF ARTIFICIAL SKY BRIGHTNESS AND UPWARD EMISSION IN ITALY FROM DMSP SATELLITE MEASUREMENTS} % ALL CAPITAL LETTERS PLEASE !!!
\author{FABIO FALCHI, PIERANTONIO CINZANO}
\institute{Dipartimento di Astronomia, Universit\`a di Padova,
vicolo dell'Osservatorio 5,\\  I-35122 Padova, Italy\\
email: falchifa@tin.it\\
email: cinzano@pd.astro.it}
\date{} % DO NOT INSERT ANY DATE HERE !!!
\end{opening}

\begin{document}

%\oddpagefooter{\sf Mem. S.A.It., Vol. ??, ??}{}{\thepage}
%\evenpagefooter{\thepage}{}{\sf Mem. S.A.It., Vol. ??, ??}
\oddpagefooter{}{}{} % LEAVE AS IT IS !
\evenpagefooter{}{}{} % LEAVE AS IT IS !
\ 
\bigskip

\begin{abstract}
We obtained the map of the zenith brightness of the night sky in Italy,
constructing a simple model. The artificial sky brightness in each site is given by the integration of the contribution produced by each unitary area of surface obtained by applying a propagation function to the upward emission in the area. This operation is a convolution of the upward emission with the propagation function. In fact, the scattering from atmospheric particles and molecules of light emitted upward by the cities spreads the light far from the sources. In pratice, we convolved the DMSP satellite night-time images of the upward light emission in Italy with a propagation function, like the Treanor Law. We used the light emission as measured  by DMSP satellite images in order to bypass  errors due to differences in the output of cities of the same population arising when using population data to estimate upward flux and in order to take into account also the contribution to the sky brightness produced from cities outside the Italian boundaries. We chose the DMSP visible band images for their negligible number of saturated pixels. Italy was cloudfree as guaranteed by inspection of IR images taken at the same time.

We also evaluated the emission versus population relationship comparing the relative emissions of a number of cities of various populations. The measured emissions increase quite linearly with the city population in the range from 1000 to 400000 inhabitants. In a preliminary analysis, more populated cities seem have a lower emission per inhabitant, so that in the range from 1000 to 3000000
inhabitants the best fitting curve to the measured emission seems to be a power law with the power $0.8$ of the city population. We did not find any dependence of city upward emission on the economic development of the area.
\end{abstract}

\section{Introduction}
The effects of artificial lighting on the brightness of the night sky in Italy were explored by Bertiau, de Graeve and Treanor in 1971 (Treanor 1971, Bertiau et al. 1973), but since then no other systematic study followed. An inquiry done by the {\it Commission for the study of light pollution} of Societ\`a Astronomica Italiana (Di Sora 1991, 1993) was limited to the situation of the night sky in the main Italian astronomical observatories. In the last quarter of century in Italy there was a great increase in the brightness of the night sky due to the increased number and efficiency of the lamps used for the outdoor artificial lighting (see e.g. Cinzano 1999b).
As a consequence, awareness of the light pollution problem by the professional and amateur astronomical communities is greatly increased. A study was needed to evaluate the new situation.
 Mappings of sky brightness for extended areas were performed by Walker (1970, 1973), Albers (1998) in USA and Berry (1976) in Canada with some simple modelling. These authors used population data of cities to estimate their upward light emission and a variety of propagation laws for light pollution in order to compute sky brightness. Recently DMSP satellite images allowed direct information on the upward light emission from almost all countries around the World (Sullivan 1989, 1991) and were used to study the increase of this flux with time (Isobe 1993; Isobe \& Hamamura 1999).
 
 In this paper we present a detailed map of artificial sky brightness in Italy. In order to bypass errors due to differences in the output of cities of same population arising when using population data to estimate upward flux and in order to take into account also the contribution to the sky brightness produced from cities outside the Italian boundaries, we constructed the map measuring directly the upward flux as detected in DSMP satellite night time images and we convolved it with a light pollution propagation law. We also studied the relation upward flux -- city population in Italy and compared it with results obtained in other countries.
 
 In section 2 we describe the satellite observations.
 The relation upward flux -- city population is presented and discussed in section 3.
 In section 4 we describe and discuss the method used to construct the map of artificial sky brightness. The map is presented in section 5 together with the map of upward emission and the map of total sky brightness. Section 6 contains our conclusions.

\section{Observations}
We studied visual images of night time Italy obtained by the  Defense Meteorological Satellite Program (DMSP) of the National Oceanic and Atmospheric Administration (NOAA). DMSP are satellites  in a low altitude (830 km) sun-synchronous polar orbit with an orbital period of 101 minutes.
Visible and infrared imagery from DMSP Operational Linescan System (OLS) instruments monitor twice a day, one in daytime and one in nightime, the distribution of clouds all over the world. At night the instrument for visible imagery is a Photo Multiplier Tube (PMT) sensitive to radiation from 410 nm to 990 nm (470-900 FWHM) with the highest sensitivity at 550-650 nm, where the most used lamps for external night-time lighting have the strongest emission: Mercury Vapour (545 nm and 575 nm), High Pressure Sodium (from 540 nm to 630 nm), Low Pressure Sodium (589 nm). The IR detector is sensitive to radiation from 10,0 $\mu m$ to 13,4 $\mu m$ (10.3-12.9 FWHM).  Every fraction of a second each satellite scans a $\sim$3 km swath extending 3000 km in east-west direction with a resolution varying from less than 3 km at the nadir to about 12 km at each end (Sullivan 1991). In our images each pixel is $2.70\pm0.02\times 2.75\pm0.03$ km wide respectively in North-South and West-East directions, a mean of five by five pixels $\sim$0.56 km wide of the high-resolution original images which are not distributed. The pixel values are currently relative values rather than absolute values because instrumental gain levels are adjusted to have a constant cloud reference brightness in different lighting conditions related to the solar and lunar illumination at the time.  Due to the limited dynamic range of the satellite detectors, the automatic gain normally saturates the most lit pixel of the largest cities. Sensitivity reaches $10^{-5}$ W $m^{-2} ~sr^{-1} ~\mu m^{-1}$ (Elvidge et al. 1997). A few images were taken with lower gain (50-24 db) and they have only a few saturated pixels. We studied two visible band images chosen for the negligible number of saturated pixels and because Italy is almost cloudfree as guaranteed by inspection of the IR image taken at the same time. The images were taken the 13th January 1997 h20:27 from satellite F12 and the 11th March 1993 h20:37 from satellite F10 from an altitude of about 800 km. 

\section{Emission vs. population relationship}
In order to evaluate the emission versus population relationship we chose a number of cities in cloudfree zones of the images, as seen in the corresponding IR images, and evaluated their emissions. To measure a city emission we summed the counts of all pixels pertinent to each city. So we obtained the relative emission of a number of cities of various populations.  The population data were taken from the 1995 Italian census provided by Istituto Italiano di Statistica (ISTAT). 

\begin{figure}
\epsfysize=6cm % fix the y-dimension and scales x-dim. to y-dim.
%\epsfxsize=8cm % fix the x-dimension and scales y-dim. to x-dim.
% Feel free to do the choice you prefer but do not exceed the x-dimension
% of the text lines
\hspace{1.5cm}\epsfbox{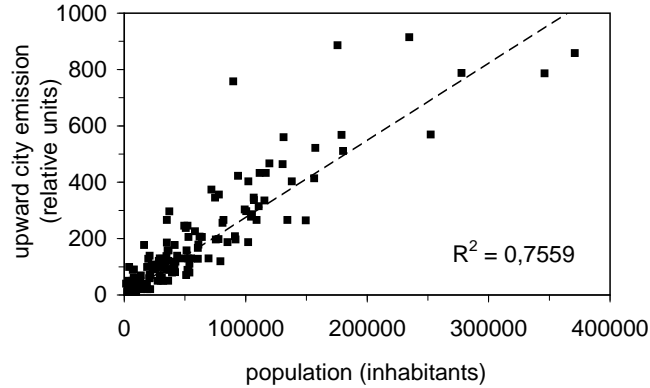} %for centering: act on hspace argument 
\caption[h]{Upward flux versus city population relationship for Italy in the range 1000 - 400000 inhabitants.}
\label{fig1}
\end{figure}

Figure \ref{fig1} show our measurements for 139 cities distributed in all Italy in the range from 1000 to 400000 inhabitants from our lower gain image. 
No sources in this plot has saturated pixels.
The dashed line shows the best fit of $u\propto P$. The upward emission increase  linearly with the population in the considered range.

\begin{figure}
\epsfysize=6cm % fix the y-dimension and scales x-dim. to y-dim.
%\epsfxsize=8cm % fix the x-dimension and scales y-dim. to x-dim.
% Feel free to do the choice you prefer but do not exceed the x-dimension
% of the text lines
\hspace{1cm}\epsfbox{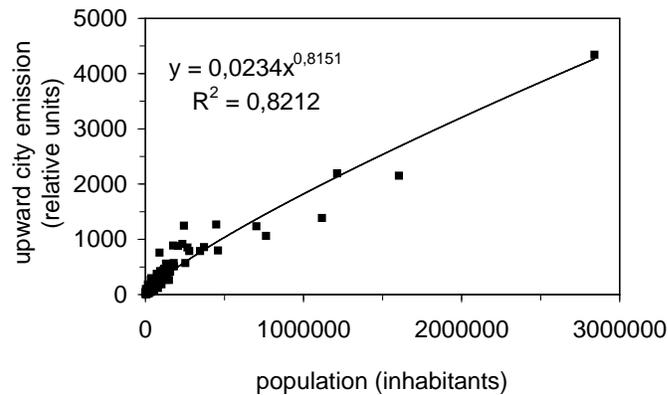} %for centering: act on hspace argument 
\caption[h]{Upward flux versus city population relationship for Italy in the range 1000 - 3000000 inhabitants.}
\label{fig1b}
\end{figure}

\begin{figure}
\epsfysize=6cm % fix the y-dimension and scales x-dim. to y-dim.
%\epsfxsize=8cm % fix the x-dimension and scales y-dim. to x-dim.
% Feel free to do the choice you prefer but do not exceed the x-dimension
% of the text lines
\hspace{2cm}\epsfbox{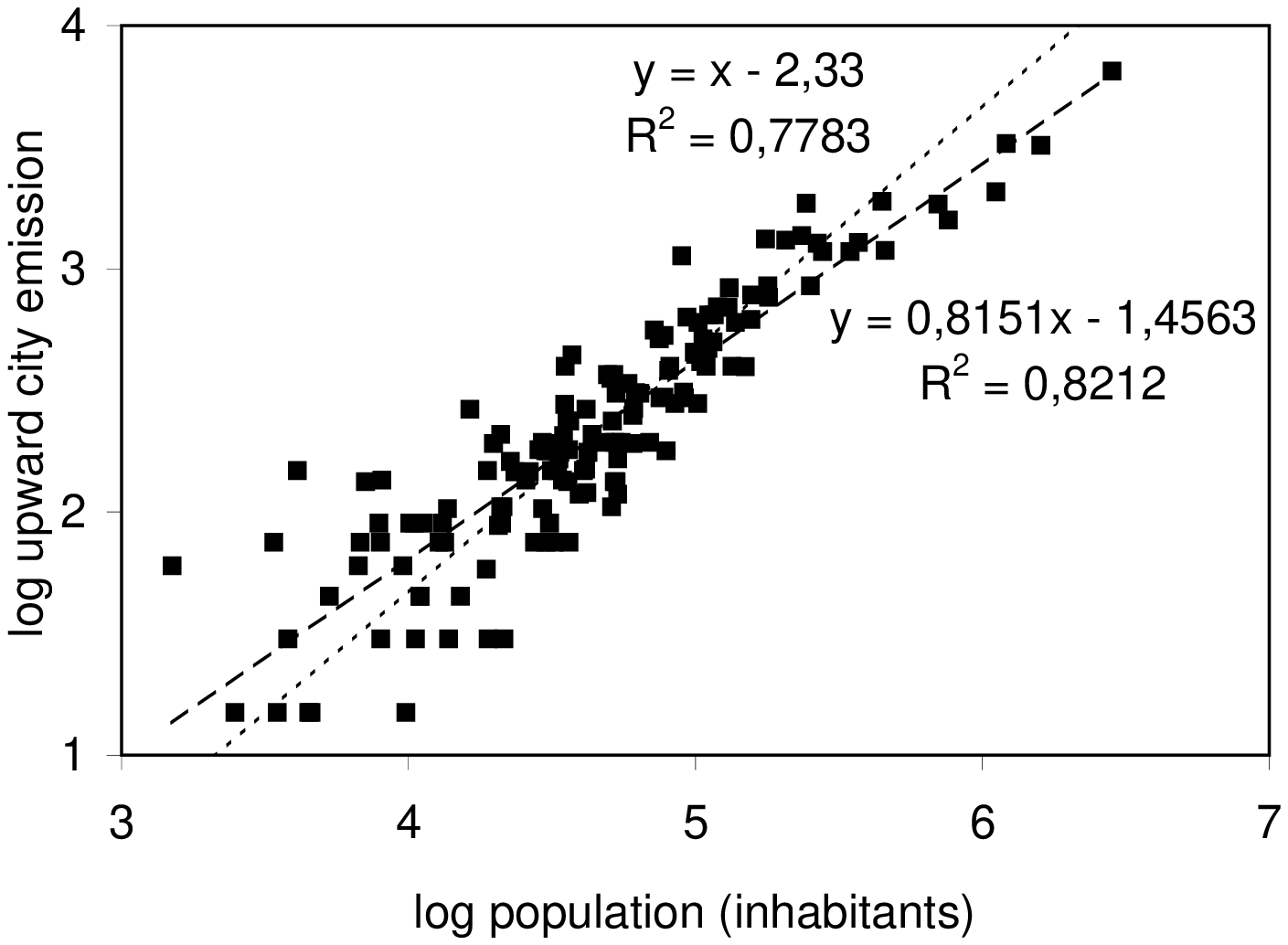} %for centering: act on hspace argument 
\caption[h]{Upward flux versus city population relationship for Italy in the range 1000 - 3000000 inhabitants in logarithmic scale.}
\label{fig1bb}
\end{figure}

 In the range 1000-3000000 inhabitants shown if figures \ref{fig1b} and \ref{fig1bb}, the best fit curve to the measured emission is a power law with the power $0.8$ of the city population $P$. These results need to be confirmed as soon as many more measurements become available.
This dependence might be due to the fact that in larger cities there are more inhabitants per unit area. In a district of high buildings there are less street lamps than in a same population district of single-family houses. This hypothesis so far was not confirmed. At the moment we cannot exclude that the exponent 0.8 be the result of the presence of saturated pixels near the center of more populated cities.

A possible source of errors  is that each pixel of the distributed images is the sum of smaller pixels and we haven't any way to check if some of them were saturated. This uncertainty will be solved only when original images will become available. 
Other main sources of error in our measurements are the uncertainties in determining the boundary of the suburbs. These errors influence mainly the measurement of the emission of the largest cities where the boundaries are less definite so that it is easier to include pixels relative to small cities in the surroundings. This causes an underestimate of the population relative to the measured emission.

\begin{figure}
\epsfysize=6cm % fix the y-dimension and scales x-dim. to y-dim.
%\epsfxsize=8cm % fix the x-dimension and scales y-dim. to x-dim.
% Feel free to do the choice you prefer but do not exceed the x-dimension
% of the text lines
\hspace{1.5cm}\epsfbox{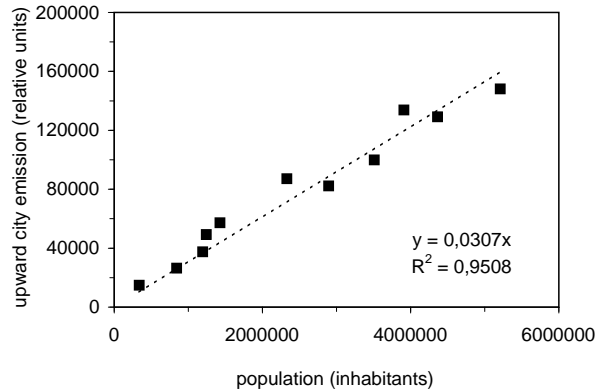} %for centering: act on hspace argument 
\caption[h]{Upward flux versus city population relationship for Italian Regions.}
\label{fig1c}
\end{figure}

Figure \ref{fig1c} shows the relation between upward emission and population for Italian Regions when the contribution of the biggest cities is subtracted. As expected, it is linear. In fact, excluding bigger cities, Regions are 
composed of a big number of sources of different populations which can be considered a statistical sample. In this case the higher gain image was used.

Even if the upward emission vs. city population function depends on the local lighting conditions, our results are in agreement with relations successifully applied or measured in other countries.
Bertiau et al. (1973) used successfully a linear model (exponent equal to 1) to model light pollution in Italy, and so did Walker (1970,1973) for studies of California skies. Garstang adopted both models with $b \propto P^{1.1}$ (Garstang 1986) and with $b \propto P$ (Garstang 1987, 1988, 1989a, 1989b, 1989c, 1991a, 1991b, 1991c, 1992, 1993) obtaining a good fit with many observations including Walker (1977) population - distance data.
Walker (1977) verified for a number of California cities the assumption of linear proportionality  for  street light emission only,  finding a general good agreement with a few departures above or below the mean depending on the industrial or residential character of the city. Walker (1977) also measured the sky illumination produced by three cities of different populations, obtaining a dependence on the power of 0.8 of their population. 
%\begin{figure}
%\epsfysize=17cm % fix the y-dimension and scales x-dim. to y-dim.
%\hspace{1.5cm}\epsfbox{tabnew1.eps} %for centering: act on hspace argument 
%\end{figure}
%\begin{figure}
%\epsfysize=16cm % fix the y-dimension and scales x-dim. to y-dim.
%\hspace{1.5cm}\epsfbox{tabnew2.eps} %for centering: act on hspace argument 
%\end{figure}
Berry (1976) fitted well observations of sky brightness in city centers in Ontario with a propagation law for light pollution based on the approach of Treanor (1973) but assuming $b \propto P^{0.5}$. Nevertheless Garstang's linear models fit well Berry (1976) observations suggesting that power of 0.5 found by him be produced by extinction of light emitted by outskirts of large cities in propagating to the center and does not depend on the upward flux versus population relationship (Garstang 1989a). 

So far, we did not find any dependence of city upward emission on the development of the area. Cities of southern Italy have the same light output as comparable size cities of northern Italy, even if the former have an  income per capita that is nearly half that of one of the latter. Bertiau et al. in the early '70 found that the city upward emission depended on its economic and commercial development, so they were forced to include in their model a development factor. There isn't any evidence in our preliminary data for the need of a similar coefficient now.

\section{Light pollution mapping technique}
We obtained the map of the zenith brightness of the night sky in Italy,
constructing a simple model. 
If $f(d)$ is a propagation law for light pollution giving the artificial sky brightness produced at a site in $(x',y')$ by an infinitesimal area $dS=dxdy$ at a distance $d$ with unitary upward emission per unit area, the total artificial sky brightness $b$ at a site is given by:
\begin{equation}
\label{int1}
b(x',y')=\int\int e(x,y) f(\sqrt{(x-x')^{2}+(y-y')^{2}})~dx ~dy
\end{equation}
where $e(x,y)$ is the upward emission per unit area from $dS$.
This expression is the convolution of the $e(x,y)$ with the function $f(\sqrt{(x-x')^{2}+(y-y')^{2}})$. So our operation consists in convolving the satellite image giving the upward emission with the propagation function. The scattering from atmospheric particles and molecules of light emitted upward by the cities spreads the light far from the sources.

In pratice, we constructed a composite image replacing the saturated pixels in the higher gain image, useful to measure accurately low population sources, with the measurements coming from the lower gain image, adequately rescaled.
We divided the surface of Italy in pixels with the same positions and dimensions as the satellite image.
We assumed each area of the country defined by a pixel be a source of light pollution with an upward emission $e_{x,y}$ proportional to the counts on the composite image and we computed the sky brightness at the center of each pixel. In this case the expression (\ref{int1}) became:
\begin{equation}
b_{i,j}=\sum_{h}\sum_{k}  e_{h,k} f(\sqrt{(x_{i}-x_{h})^{2}+(y_{j}-y_{k})^{2}})
\end{equation}

We used for $f(d)$ the
Treanor law (Treanor 1973):
\begin{equation}  
b= b_{0} I_{0} \left(
\frac{A}{d}+\frac{B}{d^{2}} \right)~~e^{-kd}  \end{equation} 
where 
$b$ is the zenith artificial sky brightness  and  $b_{0}$ is the zenith natural sky brightness at the site considered and produced by a city with upward emission $I_{0}$ placed at a distance $d$. A, B and k are constants related to the scattering component, the direct beam component and the attenuation of the city light by absorption and scattering losses. All the constants were empirically determined by Bertiau et al. (1973) to fit in the best way the data of the zenith luminance due to three different cities in Italy at various distances from them. We checked that the ratio B/A and the
coefficient k, which
depend only on the mean conditions of the atmosphere
in clean nights, were unchanged from the calibration of Bertiau et al. (1973). Differences between V-band and B-band propagation are under the fluctuations produced by changes in atmospheric conditions for clean nights (Falchi 1999).
\begin{figure}
\epsfysize=16.5cm % fix the y-dimension and scales x-dim. to y-dim.
\hspace{0.6cm}\epsfbox{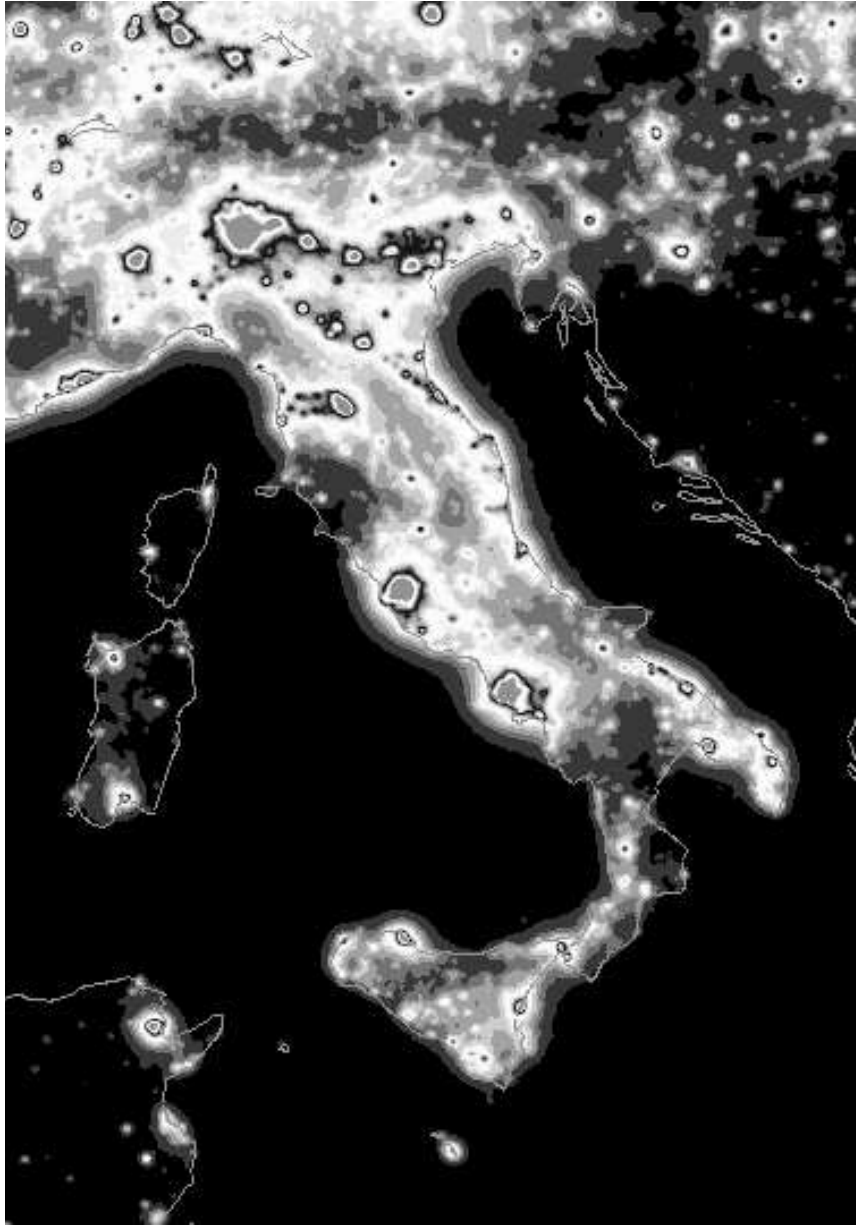} %for centering: act on hspace argument 
\epsfysize=1cm % fix the y-dimension and scales x-dim. to y-dim.
\hspace{10cm}\epsfbox{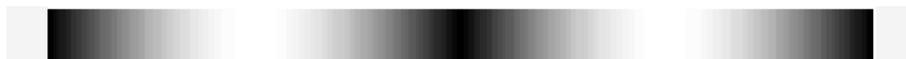} %for centering: act on hspace argument 
\caption[h]{Map of the artificial sky luminance in Italy. The counts from 0 to 100 are split into 25 levels displayed with a multiple gray scale (bottom) in order to better show the finest details of the luminance distribution. Greater counts are set to 100 for clearness purposes. Contours of Italy are just an indication.}
\label{fig2}
\end{figure}

Using directly the upward light emission as measured in the satellite images we bypassed the possible errors due to differences in the output of cities of the same population and we take into account the contribution to the sky brightness produced from cities outside the Italian boundaries. This contribution was ignored by Bertiau et al. because bordering and near lands were sparsely populated mountain regions or underdeveloped countries (Albania). This approximation was far better in the early '70 than now, because of the high increase in light emission per capita since then.
A source of error should be the possible presence of clouds over some cities; these clouds could hide or dim the light received by the satellite. To avoid this we searched for an image  without clouds over all Italian territory. We neglected curvature of the earth in our computation. This might produce an error for isolated areas but in strongly urbanized areas it is negligible. The effect of earth curvature is about 2 percent at 50 km (Garstang 1989). 
In computing the sky brightness we neglected sources outside a 200 km radius from the site. This limit was chosen to exclude overestimating the contribution from far cities whose effects at 200 km are negligible due to the earth curvature.
The emission coming from the same pixel for which we computed the artificial brightness was artificially taken away to 2.7 km distance. This was done because we want to compute the mean brightness in the area, not the exact brightness in a precise site which could be sensitive to the detailed distribution of nearest light sources. Moreover the Treanor Law breaks down at small distances, causing spurious high values due to the nearest sources.
To get the absolute values of the brightness all over Italy we need to calibrate the map with the measured brightness of the sky in some sites.  The sites should be far from bright emission sources, in order to fulfil the law used.

\section{Results}
\begin{figure}
\epsfysize=19cm % fix the y-dimension and scales x-dim. to y-dim.
\hspace{1.5cm}\epsfbox{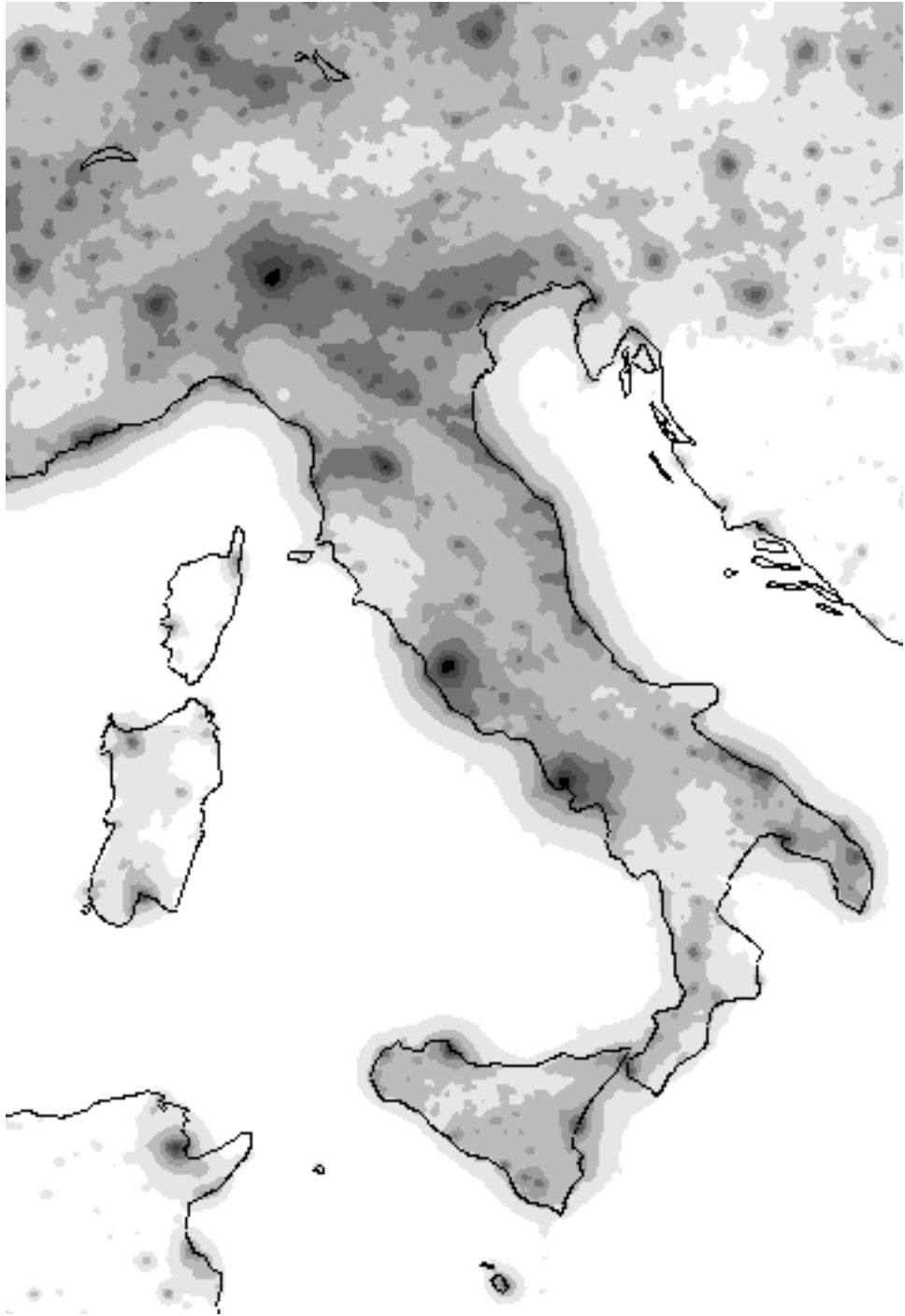} %for centering: act on hspace argument 
\caption[h]{Map of the V-band sky brightness in Italy. The eight gray levels of the image correspond for 1998 to levels of $<$0.5, 0.5-1.0, 1.0-1.5, 1.5-2.0, 2.0-2.5, 2.5-3.0, 3.0-3.5, $>$3.5 $mag/arcsec^{2}$ over the natural sky brightness. Contours of Italy are only an indication.}
\label{fig3}
\end{figure}

\begin{figure}
\epsfysize=16cm % fix the y-dimension and scales x-dim. to y-dim.
\hspace{0.2cm}\epsfbox{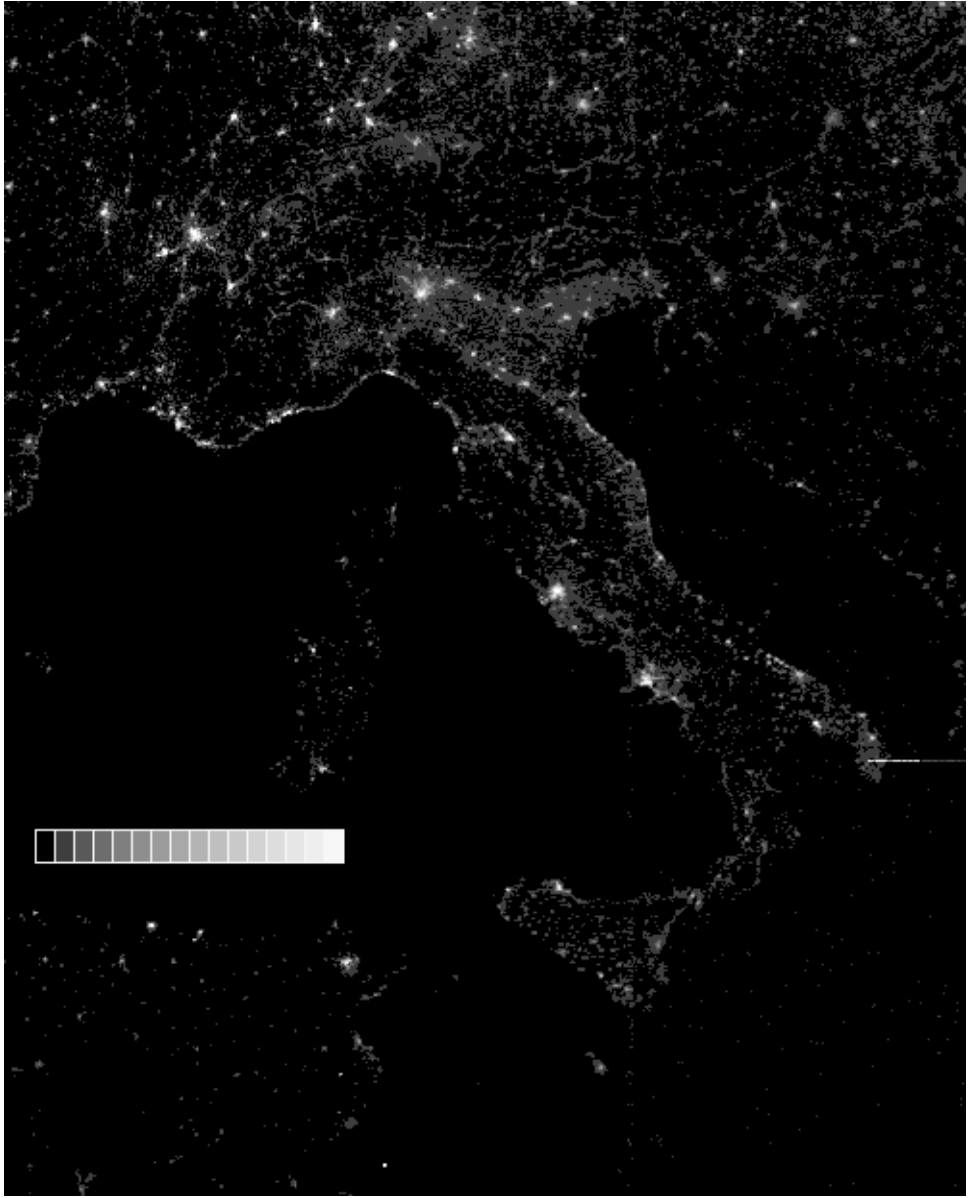} %for centering: act on hspace argument 
\caption[h]{Upward light flux in Italy. Counts go from 0 to 171 in 16 gray levels.}
\label{fig3b}
\end{figure}

The map of artificial light pollution in Italy expressed with linear scale (artificial luminance) is  in figure \ref{fig2}. We also present in figure \ref{fig3} the map of total sky brightness in Italy expressed with magnitude scale as Garstang (1986):
\begin{equation}
m_{i,j}=12.603-2.5~log(b_{i,j}+b_{nat})
\end{equation}  
where m is in V band, b is in $cd/m^{2}$ and $b_{nat}$ is the natural sky brightness for minimum solar activity assumed corresponding to 22.00 V $mag/arcsec^{2}$. Differences with Bertiau et al. (1973) due to  the increase of upward emission from time and to population changes are under study (Falchi 1999; Cinzano \& Falchi 1999). In figure \ref{fig3b} is shown a map of upward light emission as obtained directly from the satellite images.
An enlargement of figure \ref{fig2} is presented in figure \ref{fig5}.
All the maps have been obtained from the composite image including low and high gain images in order to avoid saturation and to have better sensivity to minor sources.

Satellite images are not calibrated so our upward flux measurements are only relative.
\clearpage

\begin{figure}
\epsfysize=5.4cm % fix the y-dimension and scales x-dim. to y-dim.
\hspace{-0.2cm}\epsfbox{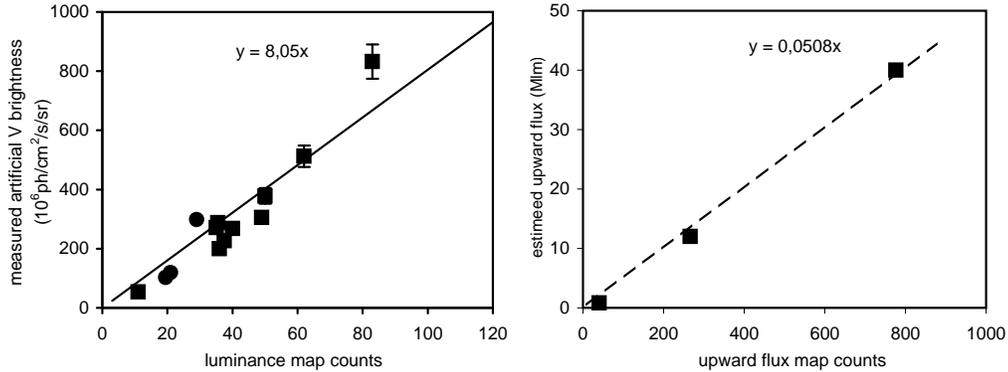} %for centering: act on hspace argument 
\caption[h]{Comparisons between maps and available data.}
\label{fig6}
\end{figure}

\begin{figure}
\epsfysize=9.5cm % fix the y-dimension and scales x-dim. to y-dim.
\hspace{1cm}\epsfbox{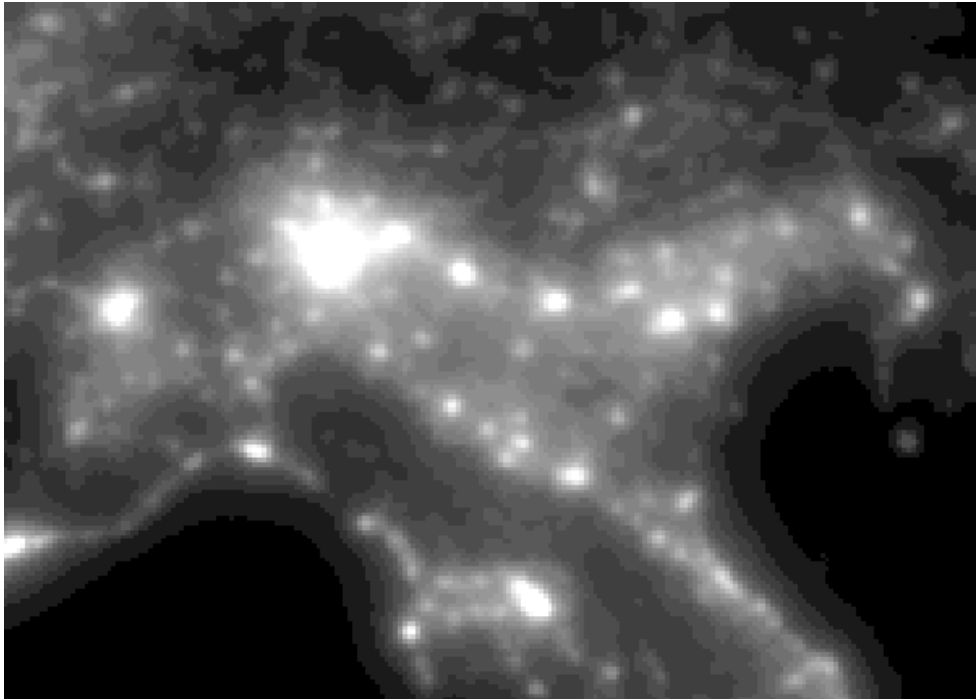} %for centering: act on hspace argument 
\caption[h]{Artificial sky luminance in North Italy. Enlargement of the map in Figure 5. Counts from 0 to 100 are displayed in 25 levels with a linear gray scale. Greater counts are set to 100 for clarity.}
\label{fig5}
\end{figure}

\noindent{We tried a preliminary calibration comparing our results with measurements of sky brightness obtained with a specific campaign. Given the growth of light pollution with time (Cinzano 1999a), we used only measurements taken in the second half of 1998. After transforming the measurements into luminances we subtracted the natural sky luminance (see Cinzano 1999b).
The comparison is presented in figure \ref{fig6} (left). Squares are our measurements. Circles are available measurements respectively at Serra La Nave, Loiano and Ekar Observatories  plotted only for comparison purposes because  their altitudes are higher than sea level assumed in our computations.}
\begin{figure}
\epsfysize=10.5cm % fix the y-dimension and scales x-dim. to y-dim.
\hspace{0.2cm}\epsfbox{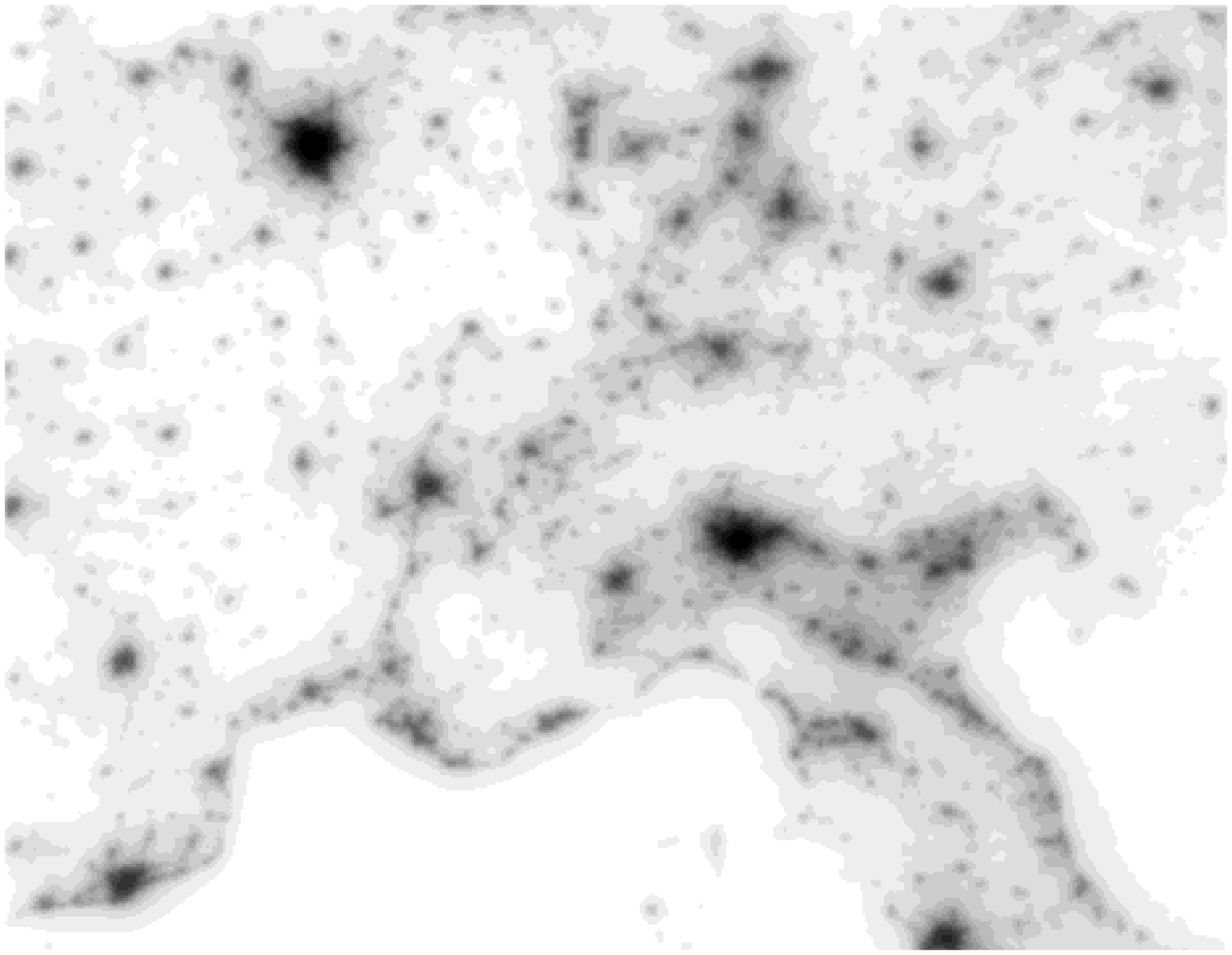} %for centering: act on hspace argument 
\caption[h]{Preliminary map of artificial sky luminance in parts of the European countries from an high gain satellite image. Due to saturation of more populated city this map is usefull only for a qualitative evaluation. The 16 gray levels are equispaced but the scale is not calibrated.}
\label{fig4}
\end{figure} 
We plan to obtain a better calibration after a larger sky brightness measurement campaign.

We also tried a comparison of our upward flux map with available data of upward flux. Results are shown in figure \ref{fig6} (right). Data refers respectively to Asiago, Treviso (from Medusa 1999, assuming public lighting 60\% of total), Padova (Favero et al. 1999). We excluded available data from the city of Torino because it has saturated pixels in all our images. At the moment the resulting calibration has to be considered highly unconfirmed.

\section{Conclusions}
We obtained the maps of the zenith artificial sky luminance and zenith total sky brightness of the night sky in Italy convolving with a light pollution propagation function the upward light emission from each unitary area as measured in DMSP satellite night-time visible band images chosen for their negligible number of saturated pixels and for the cloudfree Italy.
We are working to extend our mapping to other European countries. Figure \ref{fig4} shows a preliminary map obtained from the high gain image of 1993. Densely populated cities in this image have saturated pixels so the map is less accurate near major cities than the other maps but it could be more accurate in less populated areas. Better results will be presented in a forthcoming paper.

We studied the emission versus population relationship comparing relative emission of a number of cities of various populations. The measured emission increases quite linearly with the city population in the range from 1000 to 400000 inhabitants. More populated cities seems to have a lower emission per inhabitant, so that in the range from 1000 to 3000000
inhabitants the best fitting curve to the measured emission seems to be a power law with the power $0.8$ of the city population. At this stage we cannot confirm this relation. So far we did not find any dependence of city upward emission on the economic development of the area.

A comparison with the map of Bertiau et al. (1973) showing the changes in sky brightness from 1973 to 1997 is in preparation (Cinzano \& Falchi 1999).

\acknowledgements
We acknowledge Christopher D. Elvidge and David J. Serke of NOAA-National Geophysical Data Center for kindly providing us useful information. FF also acknowledges Dennis Geremia for his help on C programming.
We are indebted to Roy Garstang of JILA-University of Colorado for his friendly kindness in reading and refereeing this paper, for his helpful suggestions and for interesting discussions.

% References. We avoided using the \bibitem commmand since we found it is
% somewhat platform-dependent. We also avoided using the \cite{keyword}
% command since we found it cumbersome. However, if you are an expert 
% LateX user you may use the various LateX tools for the references 
% provided they give the same printout formats of the examples given here.

\end{document}